# First-principles design of stable spin qubits in monolayer MoS$_2$ with elemental defect engineering

Cailian Yu[1], Zhihua Zheng[1], Menghao Gao[1], Zhenjiang Zhao[1], Xiaolong Yao[1, 2, *]

[1]*School of Physical Science and Technology, Xinjiang Key Laboratory of Solid-State Physics and Devices, Xinjiang University, Urumqi 830017, China*

[2]*Beijing Computational Science Research Center, 100193 Beijing, China*

[*]Email: xlyao@xju.edu.cn

## ABSTRACT

Quantum information science (QIS), encompassing technologies such as quantum computing, sensing, and communication, relies on the development and manipulation of quantum bits (qubits). Recently, two-dimensional (2D) materials - characterized by their atomic thinness and external controllability - have emerged as promising candidates for qubit fabrication and manipulation at room temperature. In this study, we propose that antisite defects ($M_X$) in 2D transition metal disulfides (TMDs) can serve as tunable quantum defects with controlled positioning. Using first-principles atomic structure simulations, we identify six thermodynamically stable neutral antisite defects ($M_X$, where M = Mg, Ca, Sr, Ba, Zn, Cd; X = S) in monolayer 1H-MoS$_2$. These defects exhibit potential as spin-defected qubits with stable triplet ground states. Additionally, we demonstrate that the reduction of the bandgap leads to significant fluctuations in the absorption coefficient within the low-energy range, resulting in the optical response within the desired telecommunication band, which is advantageous for quantum communication applications. The zero-phonon line (ZPL) associated with these qubits can serve as an effective identifier. This work presents the novel, tunable approach to exploiting defects in 2D materials, opening new possibilities for the development of qubit platforms in quantum information technology.

## I. INTRODUCTION

Quantum information science has emerged as the rapidly advancing field, with defect-controlled spin qubits playing the pivotal role in enabling scalable quantum technologies. Recent experimental breakthroughs have expanded the material platforms for hosting such qubits, including $MoS_2$ transistors [1], phosphorus-doped silicon [2], and diamond NV centers [3,4]. Our study focuses on defect-engineered qubits, the paradigm that benefits from the atomic-scale precision of solid-state systems. The key priority in the field is to identify and optimize materials that balance high coherence times, tunable spin-photon interfaces, and compatibility with existing semiconductor fabrication techniques.

Atomic vacancy defects in the lattice play a crucial role in semiconductor physics, often determining the electronic, optical, and transport properties of the system. These defects are essential tuning knobs for optimizing the efficiency of solar cells, semiconductor transistors, light-emitting diodes, and catalysts. Solid-state quantum defects are ideal candidates for scalable quantum information systems that can be seamlessly integrated with conventional semiconductor electronics in three-dimensional (3D) monolithically integrated hybrid classical-quantum devices [5,6]. The prominent example is the nitrogen vacancy (NV) center defect in diamond, though controlling the localization of NV centers in bulk diamond remains the significant challenge [3,4,7-9]. Additionally, quantum defect properties in bulk crystals may not be easily tunable.

In contrast, two-dimensional (2D) semiconductors, such as transition metal dichalcogenides (TMDs), offer the promising platform to host quantum defects with tunable properties and precise positional control [10-13]. Recently, point defects in ultrathin 2D materials have attracted considerable research interest. The atomic thinness of TMDs makes them particularly susceptible to defect modulation, enabling alterations of their intrinsic properties when compared to their bulk counterparts. For example, vacancy defects can regulate the bandgap of TMDs for broad spectral detection [14,15] and enhance their electrocatalytic performance [16]. Similarly, impurity defects can not only adjust the bandgap and modify the electrical characteristics of TMDs but also alter their magnetic properties. Consequently, defect engineering - aimed at controlling defects to tune electrical, optical, magnetic, and electrocatalytic behaviors - has emerged as the central research focus. This study addresses two key aspects of defect engineering: on one hand, the suppression and repair of defects to maintain material stability and preserve original properties; and on the other, the intentional introduction of defects to achieve specific functionalities [17,18]. To date, the variety of mature technological methods have indeed been developed that can precisely control the positioning of defects in 2D materials, which is essential for fields such as quantum communication. These technologies include: Scanning Tunneling Microscopy (STM) manipulation [19,20], which allows for the creation of defects with atomic-level precision; Focused Ion Beam (FIB) milling [21,22], used for the nanoscale removal of materials; and Chemical Vapor Deposition (CVD) combined with patterned catalysts to control material growth and the introduction of defects [23,24], among many other techniques. These methods have provided effective means for the precise manipulation of defects in two-dimensional materials, advancing quantum technology, particularly in the controllable coupling of quantum bits. With further technological development, we anticipate that the control of defects in two-dimensional materials will reach the higher level of precision and controllability. Recent advances in hybrid techniques [21] now enable defect-specific manipulation with nanoscale precision. Combined with h-BN encapsulation and isotope engineering, these methods provide a viable roadmap for scalable quantum architectures in 2D materials. Such rational defect construction paves the way for new applications in energy technology and, more importantly, in quantum information technology, the next-generation information paradigm that employs quantum bits (qubits) as information carriers and processing units.

Compared to conventional bulk materials, 2D materials facilitate integration into smaller solid-state devices due to their atomically thin profiles. Defects in two-dimensional materials can exhibit a rich array of many-body physical properties, inspiring active theoretical developments in recent years. First-principles predictions of their electronic, optical, and spin properties will yield unbiased insights into designing new quantum defects and effectively controlling existing defects (e.g., spin qubits or single-photon emitter (SPEs), which are critical for emerging quantum information technology and spin photonics applications [11]. Missing atoms or atomic substitutions (point defects) in the lattice of 2D materials can host emerging quantum technologies, including single photon emitters and spin qubits [25-31]. Designing quantum defects in 2D materials based on first-principles has facilitated the discovery of rational spin qubits. These defects significantly impact the physical, electronic, chemical, and surface properties of materials [32-34]. Importantly, defects do not always adversely affect 2D materials; rather, intentional manipulation of defect types and quantities can lead to desirable features in the system. However, significant challenges persist in the rational design of defects (e.g., single-photon emitters or quantum-bit-hosting systems) for tailored quantum functionalities in 2D materials. [9,35-39]. Therefore, effective defect engineering in 2D materials is central to advancing both existing and next-generation applications.

Since the first isolation of single-layer graphene via mechanical exfoliation in 2004, this material has garnered global scientific interest due to its exceptional electronic mobility and broad-spectrum absorption properties. However, intrinsic graphene is the zero-bandgap semimetal, posing significant challenges for optoelectronic device fabrication, particularly in achieving high switching ratios [40]. In contrast, two-dimensional TMDs offer advantages such as tunable bandgaps, high photoresponsivities, and enhanced switching ratios, making them both strong competitors and valuable complements to graphene in various optical and electronic applications [4,9,11,27,35,41-44]. Recently, Y. Lee *et al.* systematically investigated $M_X$ defect families in monolayer TMDs, demonstrating their potential as solid-state quantum bits. Compared with nitrogen-vacancy (NV) centers in diamond and $C_BV_N$ defects in hexagonal boron nitride (hBN), these $M_X$ defect families exhibit ideal qubit properties and can operate at telecommunication wavelengths [10].

In this paper, we report the identification of six $MoS_2$ neutral antisite defects $M_X$

(M = Mg, Ca, Sr, Ba, Zn, Cd; X = S) within TMD systems. These findings enable the design of defective structures with various substitutional elements, thereby expanding their potential properties and applications. Our first-principles calculations demonstrate that the proposed antisites in these TMDs exhibit paramagnetic triplet states with flexible horizontal splitting, governed by positional symmetry. We have also determined their structural stability, electronic structure, and optical properties. Notably, as the main-group atomic number increases, changes in the electron localization function and absorption spectra are observed. For example, new fluctuations within the telecommunication wavelength region emerge in the absorption spectra as the main-group atomic number increases. This work provides valuable insights and directions for the design and application of these novel defect structures.

## II. CALCULATION METHODS AND MODELS

All calculations are performed using the Vienna *Ab-initio* Simulation Package (VASP) [45,46], based on density-functional theory (DFT) [47,48]. To compute the spin densities of states (DOS), we employ the projector-augmented wave (PAW) method and the plane-wave basis set. Electron-electron interactions are corrected using the Perdew-Burke-Ernzerhof (PBE) exchange-correlation functional, which is part of the generalized gradient approximation (GGA) [49]. The PBE-GGA functional assumes unscreened Coulomb interactions, appropriate for monolayers in vacuum. Substrate-induced screening (e.g., h-BN encapsulation) would require explicit treatment of dielectric effects, which is beyond the scope of this work but merits future study. The plane-wave cutoff energy of 400 eV is chosen, with the energy convergence criterion set to $10^{-6}$ eV. We conduct *ab initio* atomic molecular dynamics (AIMD) simulations to evaluate the thermal stability of the structure.

In this study, we employ the PBE functional, which is known to underestimate absolute bandgap values but reliably reproduces trends in defect-induced electronic modifications. Our analysis focuses on relative changes in bandgaps and spin properties across antisite defects, where PBE's systematic errors cancel out. Nevertheless, we emphasize that hybrid functionals (e.g., HSE) or many-body perturbation methods (e.g., GW) are required for quantitative accuracy in absolute bandgap and exciton binding energy predictions. Future studies will also address dynamic screening effects arising from charge transfer polarization, which are critical for exciton dissociation in optoelectronic devices.

All systems are simulated using the Monkhorst-Pack scheme, with the Brillouin zone sampled using the Γ-centered k-point grid. The vacuum space of 16 Å is added in the direction perpendicular to the monolayer, using the planar supercell of 6×6×1 to prevent interactions between neighboring images. Structural relaxation is performed for all systems, converging until the force acting on each ion is less than 0.02 eV/Å. The convergence criteria for the total energies of structural relaxation and self-consistency calculations are $10^{-6}$ eV.

Zero-phonon line energies between triplet states are calculated using the constrained DFT (CDFT) method. Spin-orbit coupling is implemented in VASP as the perturbative treatment, with the out-of-plane direction chosen as the quantization axis.

To calculate the optical properties, we employ the random phase approximation (RPA) method [50] to account for local field effects. The RPA method is commonly used to study the electronic response and optical properties of materials by considering electron-electron interactions beyond the single-particle energy level. In addition, we employ the real-time time-dependent density functional theory (rt-TDDFT) framework via the Octopus software package to investigate the optical properties of defective $MoS_2$. The optical absorption spectra are computed using the real-space, real-time methodology for dielectric function calculations within rt-TDDFT [51]. While these methods neglect excitonic effects and may underestimate absolute absorption intensities, they reliably capture relative trends in defect-induced modulation (e.g., absorption enhancement, spectral shifts). RPA and rt-TDDFT yield consistent trends, confirming that the observed optical changes arise from single-particle band structure modifications. Future studies requiring quantitative exciton spectroscopy will adopt BSE-level calculations.

### III. RESULTS AND DISCUSSION

#### A. Selection of structures

To computationally identify viable quantum bits of spin defects in monolayer TMDs, we first explore doping defects in monolayer $MoS_2$ [10]. Figure 1(a) illustrates the atomic structure of $M_X$ defects considered in this study within the 6×6×1 supercell of $MoS_2$, where element M replaces the X atom in the top layer (M = Be, Mg, Ca, Sr, Ba, Zn, Pd, Cd; X = S). However, it is advantageous for the doping element M to differ from the transition metal atoms constituting the host TMD to allow optical distinction of the modulated defective quantum bits from native defects and to achieve low enough concentrations for single quantum bit isolation [10]. Figure 1(b) displays the energy band structure of the spin-polarized 6×6×1 supercell of MoS$_2$, revealing the calculated direct bandgap of 1.622 eV, with the valence band maximum (VBM) and conduction band minimum (CBM) located at the Γ point. This result aligns well with the values of 1.69 eV reported by Y. Li *et al.* [52] and 1.67 eV by M. Kan *et al.* [53].

We conduct DFT calculations using the PBE functional to efficiently screen and predict the electrical properties of the materials. As indicated in Table I, the incorporation of antisite defects leads to the significant reduction in the band gap from 1.622 eV to a range between 0.283 eV and 0.467 eV. Figure S1(a)-S1(e) in the Supplemental Material depicts the alkaline-earth metal dopants from group II (Be, Mg, Ca, Sr, and Ba), showing that the Fermi energy levels shift toward the bottom of the conduction band with increasing atomic number. As shown in Figure S1(f)-S1(h) in the Supplemental Material, the transition metal dopants (Zn, Pd, Cd) demonstrate the similar trend, both groups have the same number of outermost electrons, and their increasing atomic radii influence the Fermi energy level, shifting it lower in the conduction band. Notably, the Fermi energy level in Supplemental Material Figure S1(g) approaches the top of the valence band, and for the Pd dopant, the spin-up and spin-down states completely overlap.

Figure 2(a) presents the cross-sectional schematic of the electron localization function, where the blue line indicates the position of the selected cross-section. In panels (b)-(f), the substituting atoms are alkaline-earth metals (Be, Mg, Ca, Sr, Ba). The data reveals that Be atoms exhibit the strongest electron localization, followed by Mg, while Ba shows the weakest localization. This indicates that in the $M_X$ defects of $MoS_2$, the electron localization of the doping elements decreases linearly with the increasing atomic number. In panels (g)-(i), the substitutional atoms are transition metals (Zn, Pd, Cd), and similar trends are observed. The electronic localization of Pd atoms is notably the weakest among the dopants. The diminished electronic localization of the doped elements may lead to a more dispersed electronic state throughout the structure, reducing the likelihood of localized magnetic moments. This provides the crucial explanation for why $Pd_S$ is non-magnetic (as shown in Figure S1(g) in the Supplemental Material) and contributes to the overlap of spin-up and spin-down states in the energy band diagram of $Pd_S$.

Our defect screening protocol follows established methodologies for solid-state qubit design [38]: (i) paramagnetic triplet ground state with stable charge configuration and in-gap defect levels, analogous to the $C_BV_N$ center in hBN; and (ii) spin-preserving optical transitions between triplet states that avoid defect ionization, the critical requirement for optical qubit control. Higher spin states are preferred to decouple spin from the S=1/2 paramagnetic background, enabling spin control in the zero magnetic field. Spin-state readout necessitates spin-conserving optical transitions. We calculated the density of states for various defect states (see Figure S2 in the Supplemental Material), with increasing atomic number of the main group elements, the contributions of the *p* orbitals of the doping elements to the impurity energy levels in the band gap diminish. $Pd_S$ is non-magnetic and therefore does not support the spin triplet state. In contrast, $Be_S$ exhibits the spin triplet ground state; however, its energy level lies too close to the valence band maximum, increasing the risk of defect ionization during photoexcitation. Ultimately, we identify six structural $M_X$ defects (M = Mg, Ca, Sr, Ba, Zn, Cd; X = S) that satisfy the screening criteria for further investigation.

It is essential to assess the stability of the structure after modulation in this study. To this end, we conduct thorough analysis of the thermodynamic stability using AIMD. The simulations were performed in the NVT ensemble (constant number of particles, volume, and temperature) within the 6×6×1 supercell, utilizing the time step of 3 fs and the total simulation duration of 15 ps at the temperature of 300 K. Figure S3 in the Supplemental Material illustrates the total energy from the 15 ps AIMD simulations of the $M_X$ defects in the lowest-energy structural $MoS_2$ (M = Mg, Ca, Sr, Ba, Zn, Cd; X = S). The total energies during the simulations are depicted in Supplemental Material Figure S3(a)-S3(f). The structures shown represent the front view of the supercell both initially under dynamic simulation and after 15 ps at 300 K. Notably, the energies of these five structures converge early and oscillate around the stable energy level. We observe that structures in Figure S3(a)-S3(d) in the Supplemental Material display the decreasing trend in energy with increasing main-group elements, while structures in Figure S3(e) and S3(f) in the Supplemental Material exhibit similar behavior. These findings indicate the enhancement in the stability of these structures.

## B. Defect energy levels

We find that the six identified MoS$_2$ defect types exhibit optical transitions within the telecommunication band. Complete data for the six defect structures studied are presented in Table I these calculations closely align with prior reports on Mo$_S$ defects in MoS$_2$ [10], confirming the efficacy of our simulation method for predicting defect quantum bits. Figure 3(a)-3(c) presents the lower spin triplet state of the optical excitation pathways for all three defect structures, showcasing transitions within the spin-conserving defect. The optical transitions are situated within the bandgap $E_g$, preventing single-photon ionization of the defects. The details of the remaining four structures are provided in Figure S4(a)-S4(d) in the Supplemental Material. It is important to note that the energy level splitting of the three defect energy levels in the spin-up channel is not uniform; the two highest occupied levels in the gap are doubly degenerate. This can be observed in the figures above, where the two majority spin electrons occupy the doubly degenerate $e_x$ and $e_y$ orbitals, and the optical transitions occur between the $e_{x,y}$ and $a_1$ orbitals. The quantities in parentheses in Table I estimate SOC effects for heavier elements [54,55]. Structural optimization and band analysis under SOC reveal negligible splitting magnitudes (~$10^{-3}$ eV) and no shifts in defect-state distributions or band-edge positions compared to non-SOC results (Figure 1).

## C. Defect formation energy

To understand the doping of different elements in M$_X$ in the 6×6×1 monolayer MoS$_2$ under varying synthesis environments, we computationally analyzed the defect formation energy. The defect formation energy is the crucial quantity for determining the physical realizability of the proposed defects in the host solid. The defect formation energy of the defect $X^q$ in charge state $q$ is given by the equation [56,57]:

$$E^f[X^q] = E_{tot}[X^q] - E_{tot}[pristine] - \sum_i n_i \mu_i + q[E_F + E_{VBM}^{pristine}] \qquad (1)$$

where $E_{tot}[X^q]$ and $E_{tot}[pristine]$ are the total energies of the supercell with and without the defect $X^q$, respectively. Here, $n_i$ denotes the number of atoms of type $i$ added to the supercell ($n_i > 0$) or removed ($n_i < 0$), and $\mu_i$ represents the chemical potentials of these species, which will be elaborated upon in the next paragraph. $E_{VBM}^{pristine}$ is the valence band maximum (VBM) of the pristine supercell, and $E_F$ is the energy level of the supercell relative to the VBM. The formation energies calculated using this equation are highly sensitive to the defect concentration within the 6×6×1 supercell of $MoS_2$.

The chemical potential is contingent on the experimental growth conditions, which may vary from Mo-rich to S-rich. Therefore, we can impose explicit constraints on the chemical potential in our calculations, which is invaluable for interpreting the results. The Mo chemical potential $\mu_{Mo}$ has the upper limit; under extreme Mo-rich conditions, $\mu_{Mo} = \mu_{Mo}^{bulk}$. Similarly, under extreme S-rich conditions, $\mu_S = \mu_S^{bulk}$ provides the upper limit for $\mu_S$. In addition to these upper limits, we can determine the lower bounds using expression (2). Furthermore, the lower limits can be more precisely established using Eqs (3a)–(3b) [58-60]:

$$\mu_{Mo} + 2\mu_S = E_{tot}[MoS_2] \qquad (2)$$

where $E_{tot}[MoS_2]$ is the total energy of the bulk $MoS_2$ unit cell. The chemical potentials of these elements must satisfy the following relations:

$$\mu_{Mo} \leq \mu_{Mo}^{bulk} \tag{3a}$$

$$\mu_{S} \leq \mu_{S}^{bulk} \tag{3b}$$

The equation for the heat of formation of MoS$_2$ can be expressed as:

$$\Delta = E_{tot} - \mu_{Mo}^{bulk} - 2\mu_{S}^{bulk} \tag{4}$$

This value must be negative for stabilized compounds, our calculated value is -2.052 eV, indicating the validity of Eqs (5a)-(5d):

$$\mu_{Mo}^{min} = \mu_{Mo}^{bulk} + \Delta \tag{5a}$$

$$\mu_{S}^{min} = \mu_{S}^{bulk} + \Delta \tag{5b}$$

The constraints for the chemical potentials become:

$$\mu_{Mo}^{bulk} + \Delta \leq \mu_{Mo} \leq \mu_{Mo}^{bulk} \tag{5c}$$

$$\mu_{S}^{bulk} + \Delta \leq \mu_{S} \leq \mu_{S}^{bulk} \tag{5d}$$

For impurities, it is essential to consider the corresponding elemental chemical potential $\mu_X$. This potential has a lower limit of minus infinity, indicating the complete absence of impurities in the growth environment. The upper limit of the chemical potential for impurities is given by the energy of the elemental bulk phase:

$$\mu_X \leq \mu_X^{bulk} \tag{6}$$

Under Mo-rich conditions for MoS$_2$, the defect formation energies are lower than those under S-rich conditions (see Figure 4 and S5 in the Supplemental Material). The formation energies of the M$_X$ defects for the remaining four structures are presented in Supplemental Material Figure S6. Additionally, the relationship $E_f[M_X] < E_f[M_I] + E_f[V_X]$ indicates that the formation of M$_X$ defects is energetically favorable. The calculated formation energies for $Mo_S$ defects in $MoS_2$ are 3.872 eV and 7.971 eV under Mo-rich and S-rich conditions, respectively. These values align more closely with the 3.211 eV and 6.982 eV reported by Y. Lee *et al*. [10], suggesting that our choice of computational methodology and parameters is reasonable and yields reliable results. The formation energy of $M_X$ (M = Mg, Ca, Sr, Ba, Zn, Cd; X = S) in $MoS_2$ is somewhat lower compared to $Mo_S$ in $MoS_2$. Both figures indicate that the formation energy of $M_X$ with the alkali metal doping element is lower than that of $M_X$ with the transition metal doping element, suggesting that $M_X$ defects can be easily created. Based on this formation energy, we can create $M_X$ defects by annealing systems with pre-existing $M_I$ and $V_X$ defects.

Similar formation energy diagrams for $M_X$ defects in the family (see Figure S6 in the Supplemental Material) further support the feasibility of $M_X$ defect creation. It is important to note that the doped M must differ from the transition metal atoms that comprise the host TMDs to distinguish the intentionally created defects. Given that $V_X$ is prevalent in TMDs, $M_X$ defects are expected to form near additional $M_I$ following annealing. A variety of mature techniques - including STM manipulation, FIB milling, and catalyst-patterned CVD - enable precise defect positioning in 2D materials. While multi-qubit coupling remains the long-term challenge, the rational design of individual qubits (as demonstrated here) provides essential groundwork for scalable quantum architectures [4]. Therefore, it is speculated that M atoms can be introduced through ion implantation or STM lithography [19]. Conversely, $Ca_S$ defects can be obtained via other experimental methods [21,23], as their formation energy is lower than that of the majority of reported defect formation energies, indicating the favorable formation process.

### D. Zero-phonon line emission

To investigate the transitions between the triplet ground state ($^3A_2$) and the triplet excited state ($^3E$), we perform CDFT calculations, in which the occupation of the Kohn-Sham orbitals is fixed to the ideal configuration. This approach, as previously reported [61], facilitates the analysis of the relevant transition processes. The defect level occupancy of the excited state is set to $e_x^{0.5}e_y^{0.5}a_1^1$. In combination with PBE calculations, this method enabled us to evaluate the zero-phonon line (ZPL) of $M_X$ defects in $MoS_2$. Figure 5 presents the schematic representation of the internal energy transitions between the triplet ground and excited states, while the corresponding energy values are provided in Table I. The ZPL energies for the $M_X$ defect family are generally on the order of 1 eV, placing them near or even within the telecommunication band region (for example, 0.961 eV for $Ba_S$ in $MoS_2$). As discussed in subsequent sections, the two-dimensional host environment modulates the ZPL energy through various structural adjustments, allowing the system to respond differently across distinct spectral ranges.

### E. Optical properties

Beyond examining the geometric and electronic structures, we also investigate the optical properties of defect-containing $MoS_2$ systems by evaluating the frequency-dependent dielectric functions. In this section, we compare the optical characteristics of pristine $MoS_2$ with those of defect-containing structures.

Figure 6(a)–6(c) show the light absorption spectra of the 3×3×1, 4×4×1, and 5×5×1 supercell defect systems, respectively. Smaller supercells (3×3×1 to 5×5×1) are employed to (i) validate spectral convergence, (ii) isolate defect concentration effects, and (iii) confirm that key optical features are intrinsic to the defect physics rather than finite-size artifacts. As shown in Figure 6(a), the presence of defects leads to the considerable enhancement in light absorption within the energy range of 0.740–0.984 eV, accompanied by more pronounced fluctuations than those observed in the pristine $MoS_2$ spectrum. Notably, the $Cd_S$ defect structure exhibits the most prominent response in this energy interval. Similar trend is observed in Figure 6(b), where the $Ba_S$ defect structure demonstrates the strongest absorption enhancement in the same range. Figure 6(c) confirms the comparable pattern to those seen in Figure 6(a) and 6(b). rt-TDDFT calculations reveal significant absorption enhancement (0.740-0.984 eV) in defective $MoS_2$, with $Cd_S$ and $Zn_S$ defects showing the most and least pronounced improvements, respectively (Fig. S7). These trends align with RPA results (Fig. 6(a)), underscoring the robustness of our conclusions across methodologies.

Overall, the absorption enhancement increases gradually with the introduction of main-group elements (Group II: Mg, Ca, Sr, Ba; transition metals: Zn, Cd), especially around 1.00 eV. Moreover, the magnitude of this enhancement becomes more pronounced as the main-group atomic number increases, particularly within the 0.740-0.98 4eV region. Although the extent of enhancement diminishes slightly as the supercell size increases from 3×3×1 to 4×4×1 and 5×5×1, the general upward trend in absorption persists. Based on these observations, we anticipate that the 6×6×1 defect-containing supercells will also exhibit enhanced absorption in the telecommunication wavelength region, and this enhancement is likely to become more pronounced with increasing main-group atomic number.

## IV. CONCLUSIONS

In summary, we employ first-principles simulations to investigate $MoS_2$ structures doped with various elements. Our findings identify six defect configurations, $M_X$ (M = Mg, Ca, Sr, Ba, Zn, Cd; X = S), that exhibit stable spin triplet states in their ground states, indicating their potential as spin-defect quantum bit candidates. DOS analyses elucidate the elemental contributions to these spin triplet states. While pristine $MoS_2$ has the direct band gap of 1.622 eV, the band gaps of the six defective structures decrease to between 0.467 eV and 0.283 eV. Additionally, the formation energies of these defects are significantly lower than those reported for other defect structures, suggesting the higher theoretical feasibility of formation. Optical property calculations reveal that, compared to pristine $MoS_2$, the defective structures exhibit substantially higher optical absorption coefficients in the telecommunication band region, highlighting the significant impact of defect modulation on the material's optical properties. The successful modulation of spin triplet states underscores their promising applications in quantum communication and offers new insights for selecting 2D materials to construct quantum bits. Our electronic structure and optical property analyses provide a deeper understanding of the effects of these defects on $MoS_2$, contributing to the broader knowledge of similar materials.

## ACKNOWLEDGMENTS

This work is sponsored by the Natural Science Foundation of Xinjiang Uygur Autonomous Region (Grant No. 2022D01C689, 2023D01D03, and 2022D01C48).

TABLE I. Summary of computed defect properties in TMDs. Bold numbers denote occupied states, all values in the table were theoretically estimated in this work, and the numbers between brackets correspond to the spin orbit coupling (SOC) results.

| Defect | $E_g$ (eV) | Defect levels (eV) | | | | | | $E_{ZPL}$ |
|---|---|---|---|---|---|---|---|---|
| | | Spin up | | | Spin down | | | |
| | | $e_x$ | $e_y$ | $a_1$ | $e_x$ | $e_y$ | $a_1$ | |
| Mo$_S$ | 0.149 | **0.013** | **0.013** | **0.560** | 0.683 | 0.683 | 1.210 | 0.225 |
| Be$_S$ | 0.462 | **0.226** | **0.226** | **1.784** | 0.695 | 0.695 | 1.932 | 1.457 |
| Mg$_S$ | 0.467 | **0.747** | **0.747** | **1.608** | 1.221 | 1.221 | 1.673 | 0.573 |
| Ca$_S$ | 0.465 | 0.885 | 0.885 | -0.249 | **1.357** | **1.357** | **0.181** | 1.021 |
| Sr$_S$ | 0.461 | 0.992 | 0.992 | -0.188 | **1.460** | **1.460** | **0.100** | 1.055 |
| Ba$_S$ | 0.439 | 1.055(1.144) | 1.055(1.149) | 0.034(0.113) | **1.502(1.498)** | **1.502(1.508)** | **0.284(0.286)** | 0.961 |
| Zn$_S$ | 0.439 | **0.595(0.594)** | **0.595(0.600)** | **1.385(1.384)** | 1.042(1.041) | 1.042(1.047) | 1.557(1.555) | 0.467 |
| Pd$_S$ | 1.374 | 1.432 | 1.432 | 0.058 | 1.432 | 1.432 | 0.058 | 1.376 |
| Cd$_S$ | 0.283 | **0.800(0.805)** | **0.800(0.810)** | **1.093(1.096)** | 1.244(1.233) | 1.244(1.241) | 1.263(1.277) | 0.157 |

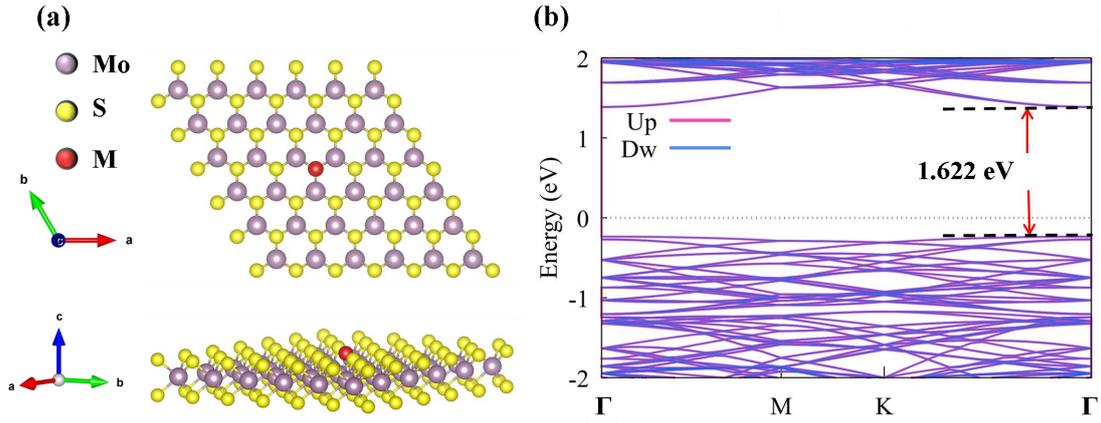

FIG. 1. M$_X$ defect geometry and primordial structural energy bands in the 6×6×1 supercell of MoS$_2$ monolayers. (a) Top view (top) and side view (bottom) of the ground-state defect geometry, yellow and purple spheres represent sulfur (S) and molybdenum (Mo), respectively, while the red sphere indicates the elemental M doped in the M$_X$ defect (M = Be, Mg, Ca, Sr, Ba, Zn, Pd, Cd; X = S). (b) Spin-polarized energy band structure for the 6×6×1 monolayer MoS$_2$. The red line corresponds to spin-up states, and the blue line corresponds to spin-down states, with spin-up and spin-down bands coinciding exactly.

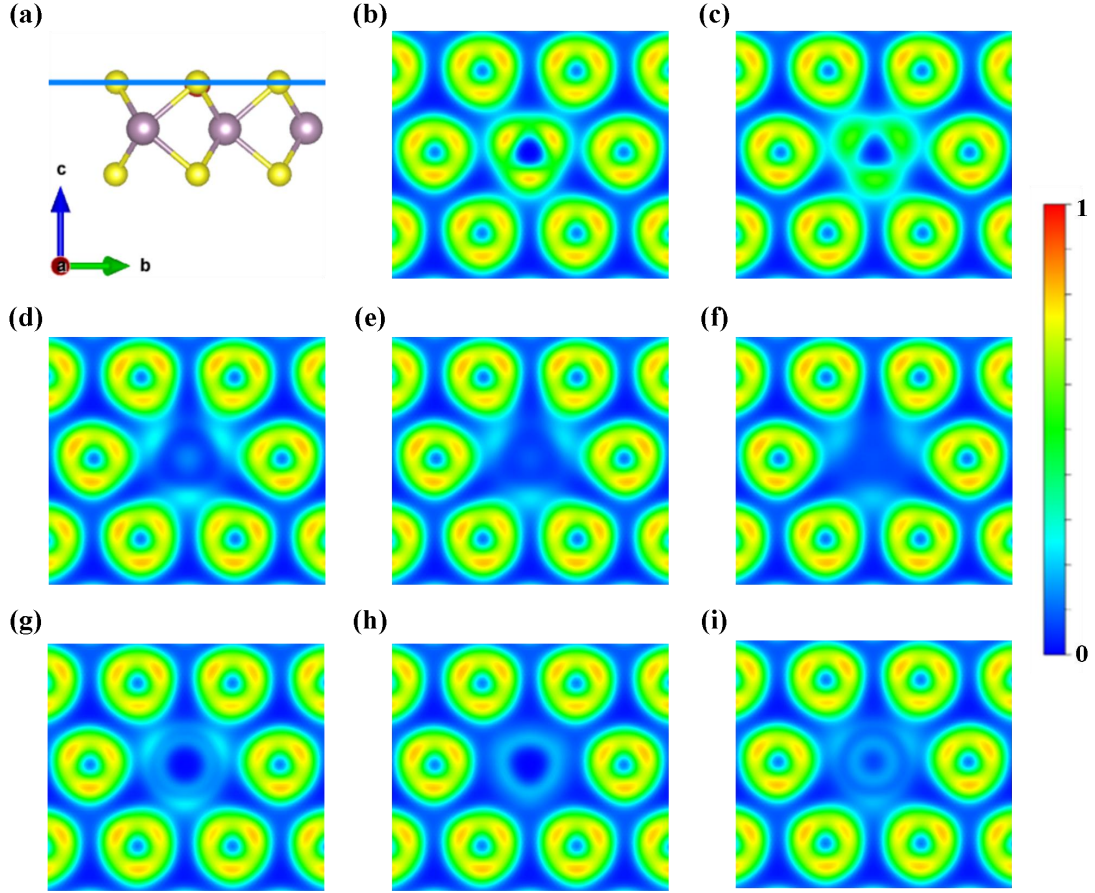

FIG. 2. (a) The schematic of $M_X$ electron localization function (ELF) cross-sections in MoS$_2$ at the isosurface level of 0.264. Panels (b)–(f) depict dopant elements from the alkaline earth metals (M = Be, Mg, Ca, Sr, Ba), while panels (g)–(i) illustrate dopant elements from the transition metals (M = Zn, Pd, Cd). The color scale on the right ranges from blue to red, representing the ELF values from 0 to 1.

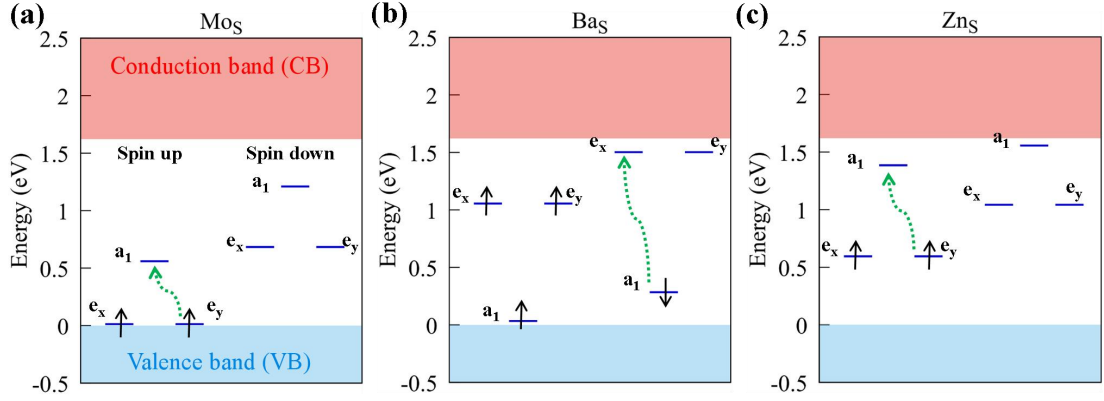

FIG. 3. Electronic structure of $M_X$ defects in $MoS_2$. Panels (a)–(c) display the energy levels of defect states for Mo, Ba, and Zn substituting S atoms, respectively. Green arrows denote spin-conserving intra-defect optical transitions. Detailed physical parameters of the $M_X$ defects are summarized in Table 1. Defect formation energy for other $M_X$ defects in $MoS_2$ are provided in Figure S3.

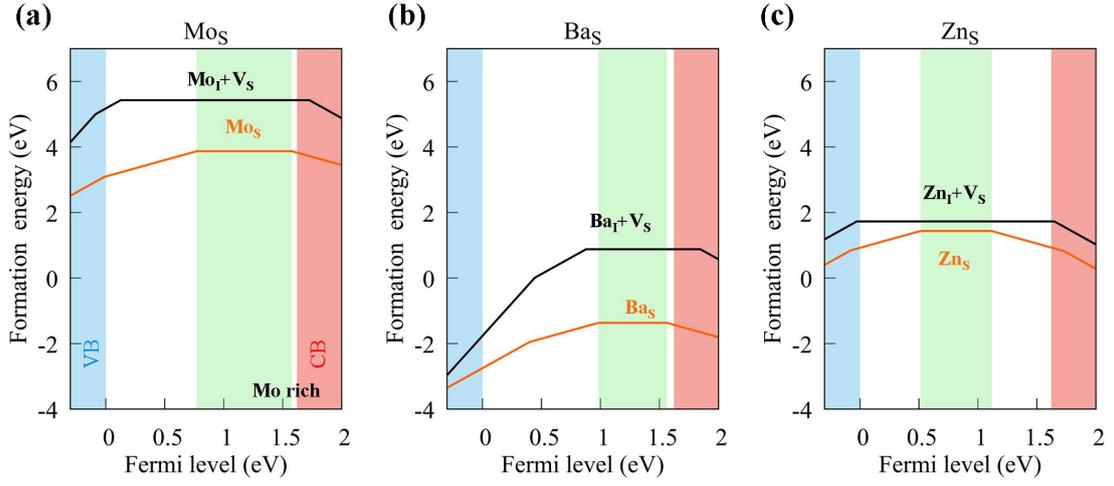

FIG. 4. Defect formation energy plots for $MoS_2$ under Mo-rich conditions: (a) $Mo_S$ in $MoS_2$, (b) $Ba_S$ in $MoS_2$, and (c) $Zn_S$ in $MoS_2$. The black dashed line labeled $M_I + V_S$ represents the sum of the formation energies of two independent defects, $M_I$ and $V_S$. The green shaded area indicates the stability range of $Mo_S^0$, $Ba_S^0$, and $Zn_S^0$. Defect formation energy maps for other $M_X$ defects in $MoS_2$ are provided in Figure S5.

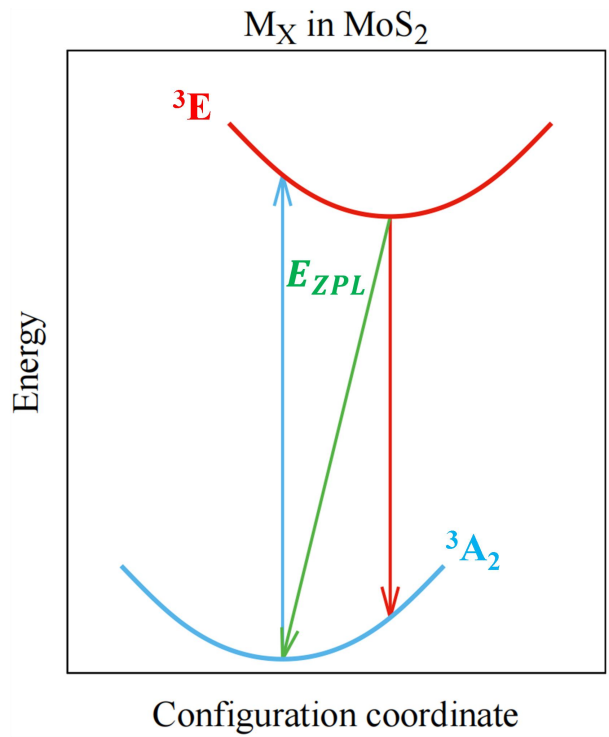

FIG. 5. Configurational coordinate diagram illustrating the zero-phonon line (ZPL) emission mechanism. The diagram depicts the defect qubits of $M_X$ in $MoS_2$, with the ground state labeled as $^3A_2$ and the excited state as $^3E$.

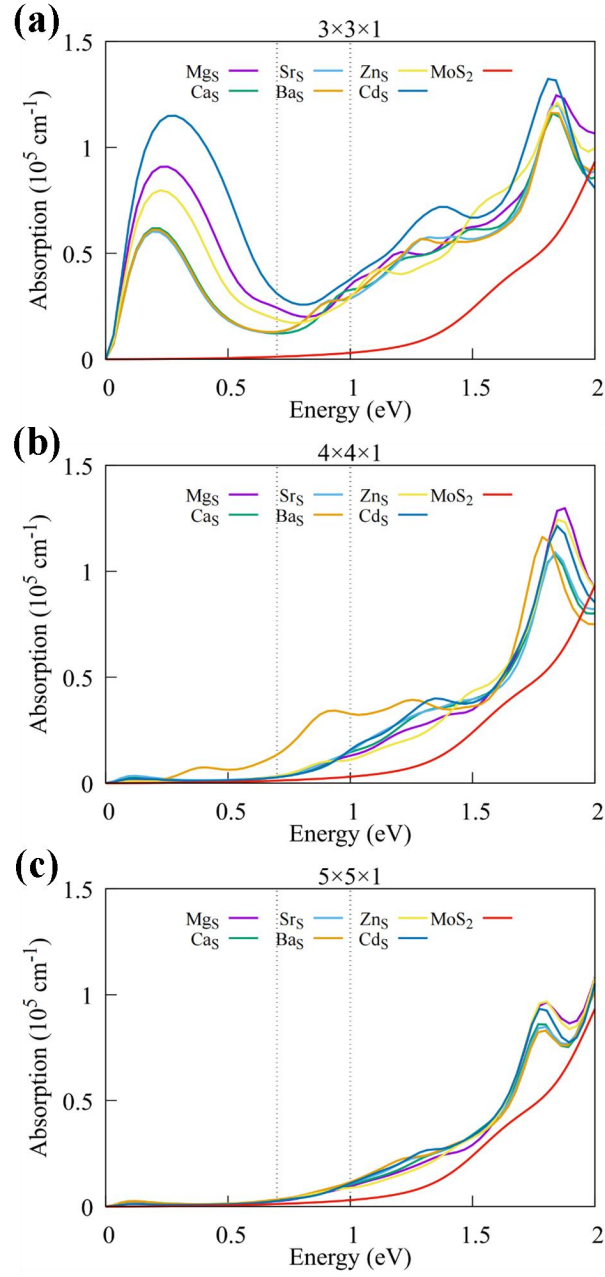

FIG. 6. Absorption spectra of monolayer MoS$_2$ with M$_X$ defects: (a) 3×3×1 supercell, (b) 4×4×1 supercell, and (c) 5×5×1 supercell. The red line represents the absorption spectrum of pristine MoS$_2$. Absorption curves are averaged over the *x* and *y* crystallographic directions. The region between the gray lines (0.740–0.984 eV) highlights defect-induced absorption features arising from mid-gap states critical for telecom-band optical transitions.

# Supplementary Material

# First-principles design of stable spin qubits in monolayer MoS$_2$ with elemental defect engineering

Cailian Yu[1], Zhihua Zheng[1], Menghao Gao[1], Zhenjiang Zhao[1], Xiaolong Yao[1,2,*]

*[1]School of Physical Science and Technology, Xinjiang Key Laboratory of Solid-State Physics and Devices, Xinjiang University, Urumqi 830017, China*

*[2]Beijing Computational Science Research Center, 100193 Beijing, China*

*Email: xlyao@xju.edu.cn

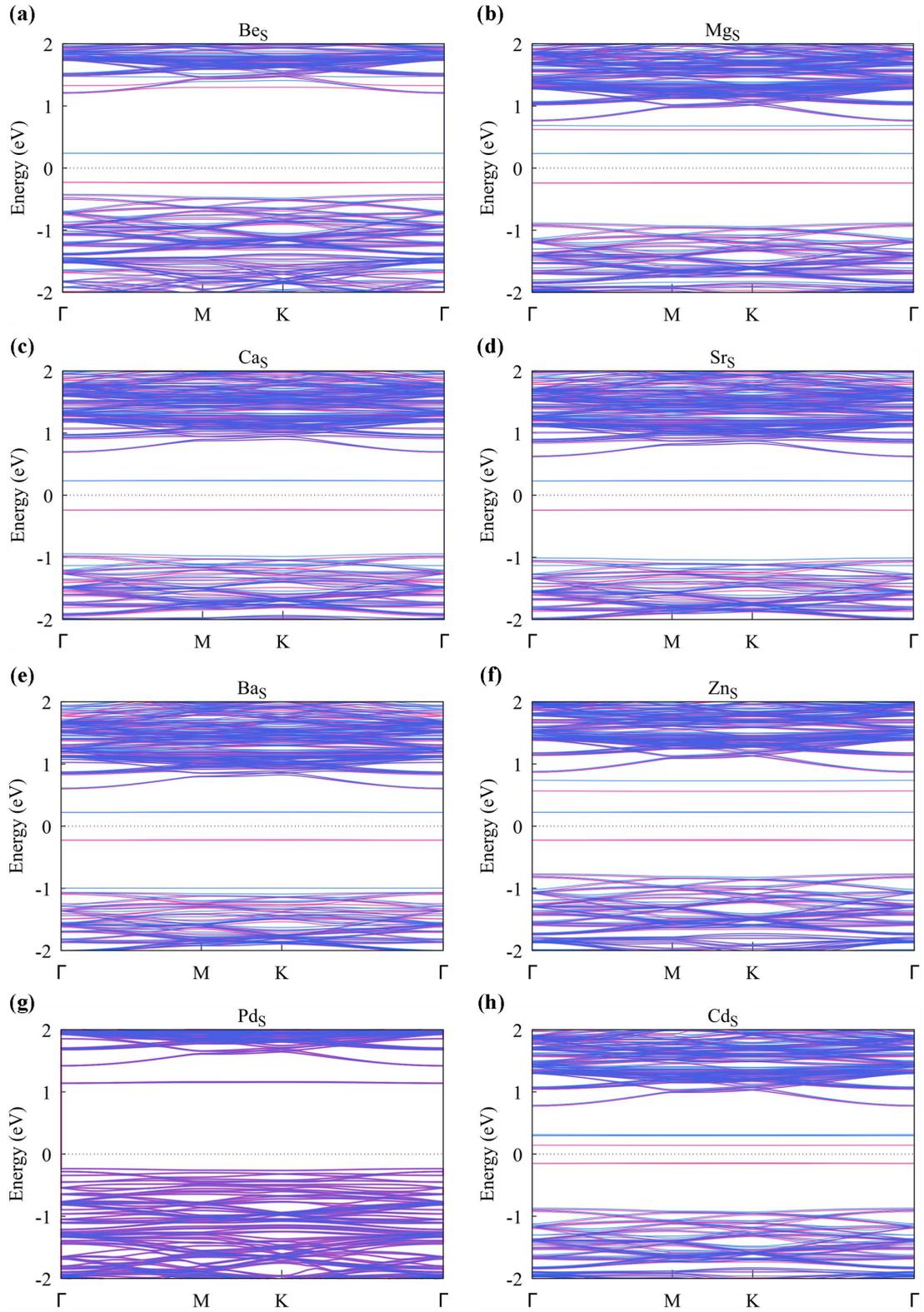

FIG. S1. Energy band structures of $M_X$ defects (M = Be, Mg, Ca, Sr, Ba, Zn, Pd, Cd; X = S) in 6×6×1 supercells of $MoS_2$. Panels (a)–(h) correspond to each dopant element, respectively. The Fermi energy is set at 0 eV.

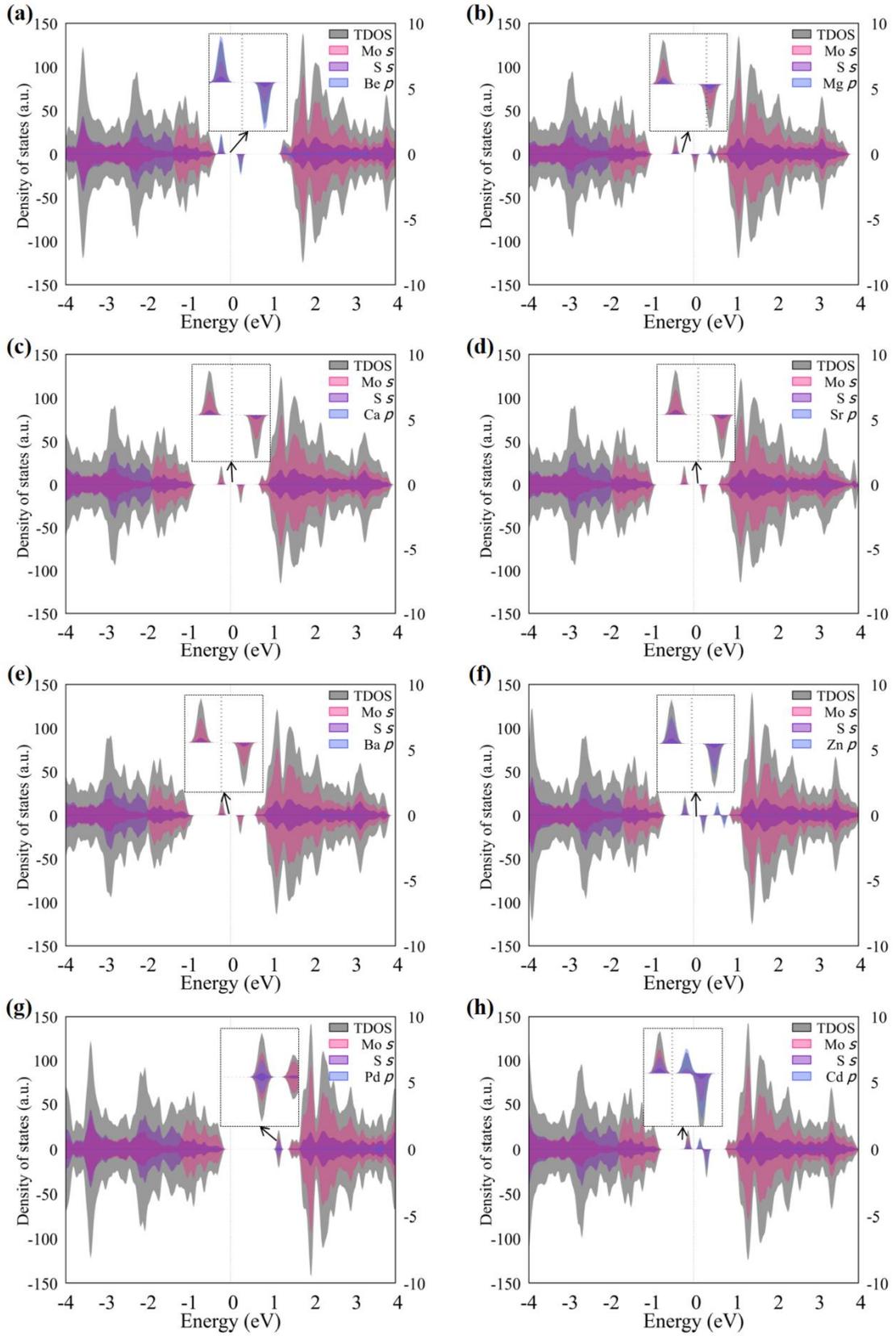

FIG. S2. Identification of spin triplet states in the ground state. Panels (a)–(h) present the density of states (DOS) for various doped defects in MoS$_2$, calculated as the function of energy relative to the Fermi level, without employing hybridization

functions for rapid screening. Notably, panel (g) indicates that spin triplet states are absent in the ground state for $Pd_S$. In contrast, defect structures other than $Be_S$ exhibit spin triplet states in the ground state, characterized by well-separated defect energy levels.

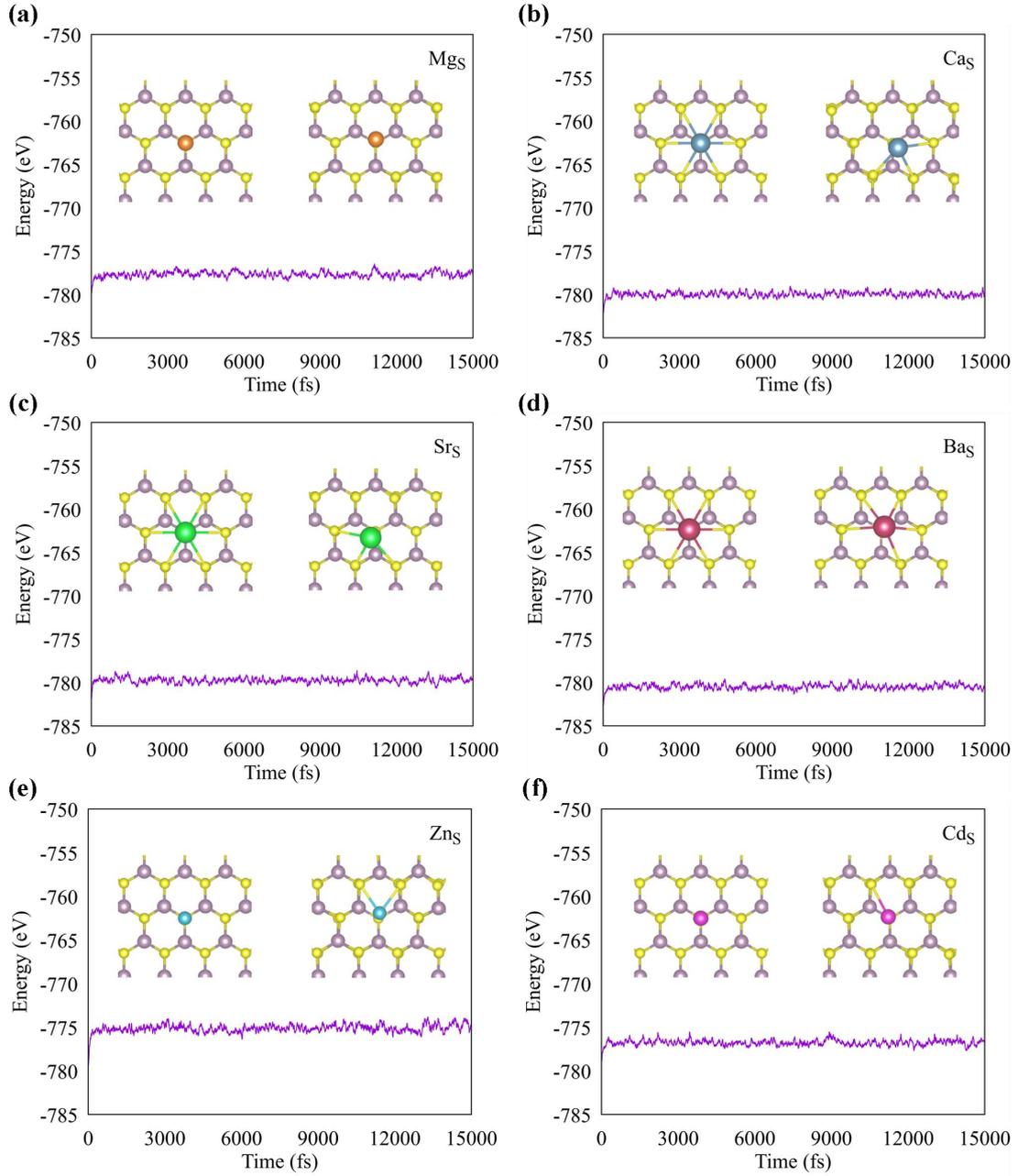

FIG. S3. Time evolution of the total energy during *ab initio* molecular dynamics simulations for six $M_X$ defects (M = Mg, Ca, Sr, Ba, Zn, Cd; X = S) in the 6×6×1 supercell of $MoS_2$ at 300 K. Panels (a)–(f) correspond to each defect, respectively. Insets depict the atomic structures at the initial and final simulation times.

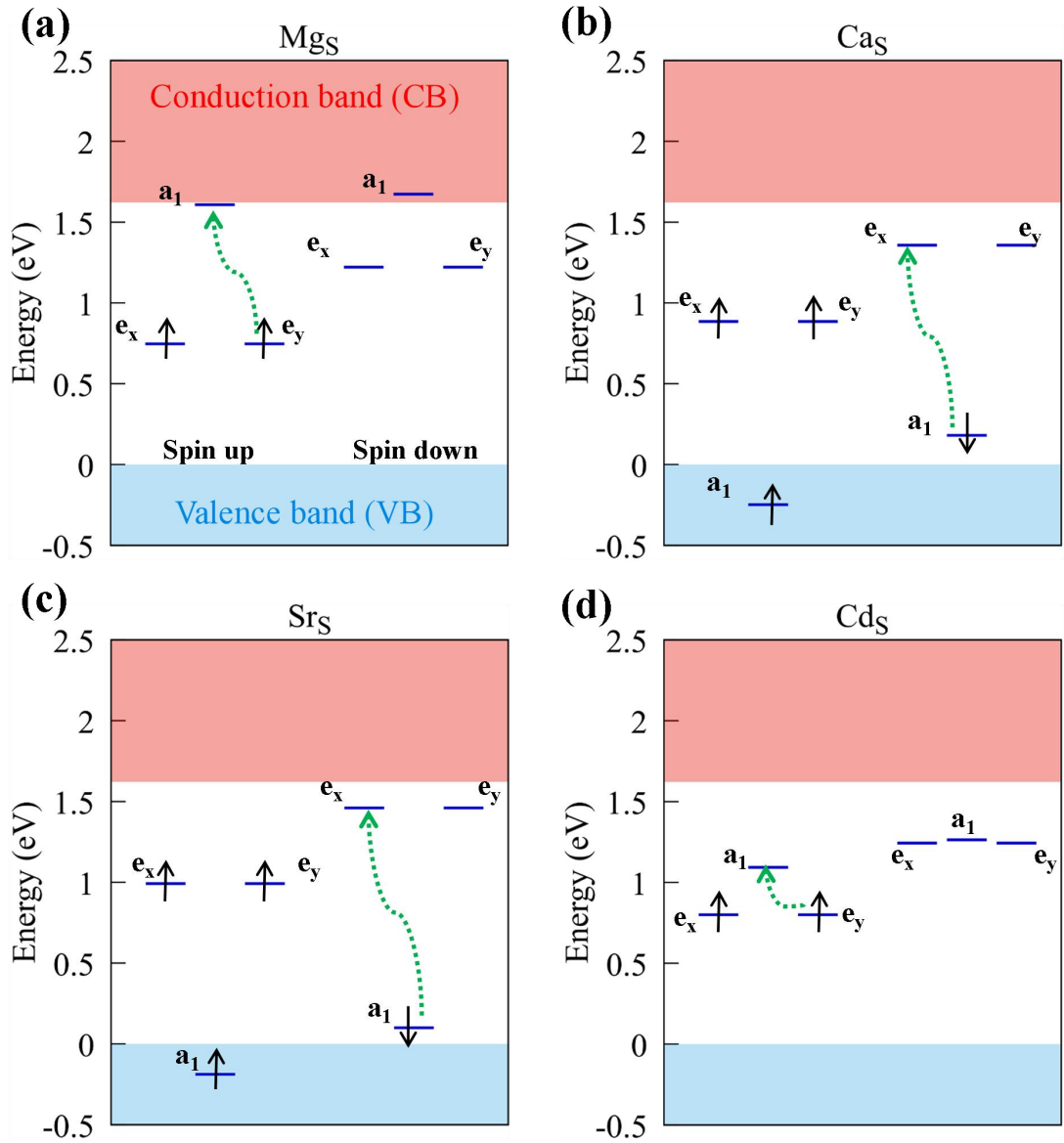

FIG. S4. Electronic structure diagrams of $M_X$ defects (M = Mg, Ca, Sr, Cd; X = S) in $MoS_2$. Panels (a)–(d) correspond to each dopant element, respectively. The energy levels of the defect sites are depicted as the function of electron occupancy.

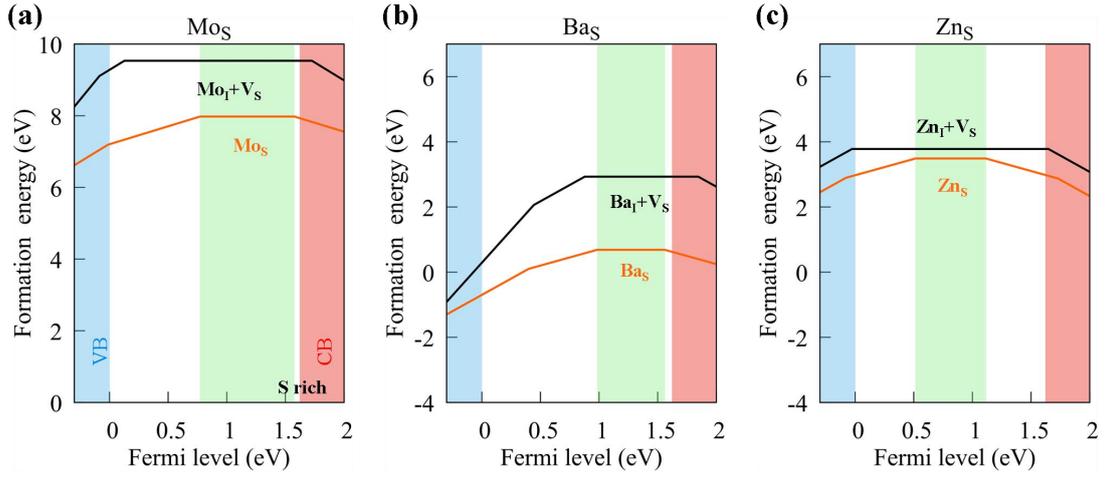

FIG. S5. Defect formation energies under varying conditions. Panels (a)–(c) depict the defect formation energies for (a) $Mo_S$ in monolayer $MoS_2$, (b) $Ba_S$ in monolayer $MoS_2$, and (c) $Zn_S$ in monolayer $MoS_2$ under different conditions, as referenced in Figure 4(a–c). Compared to the Mo-rich condition shown in Fig. 4(a), the formation energy of $Mo_S$ in $MoS_2$ is higher under the S-rich condition. Similarly, as illustrated in Figs. 4(b)-4(c), the formation energies of $Ba_S$ and $Zn_S$ in $MoS_2$ are also higher under the S-rich condition than in the Mo-rich condition. This trend persists across the entire $M_X$ defect family.

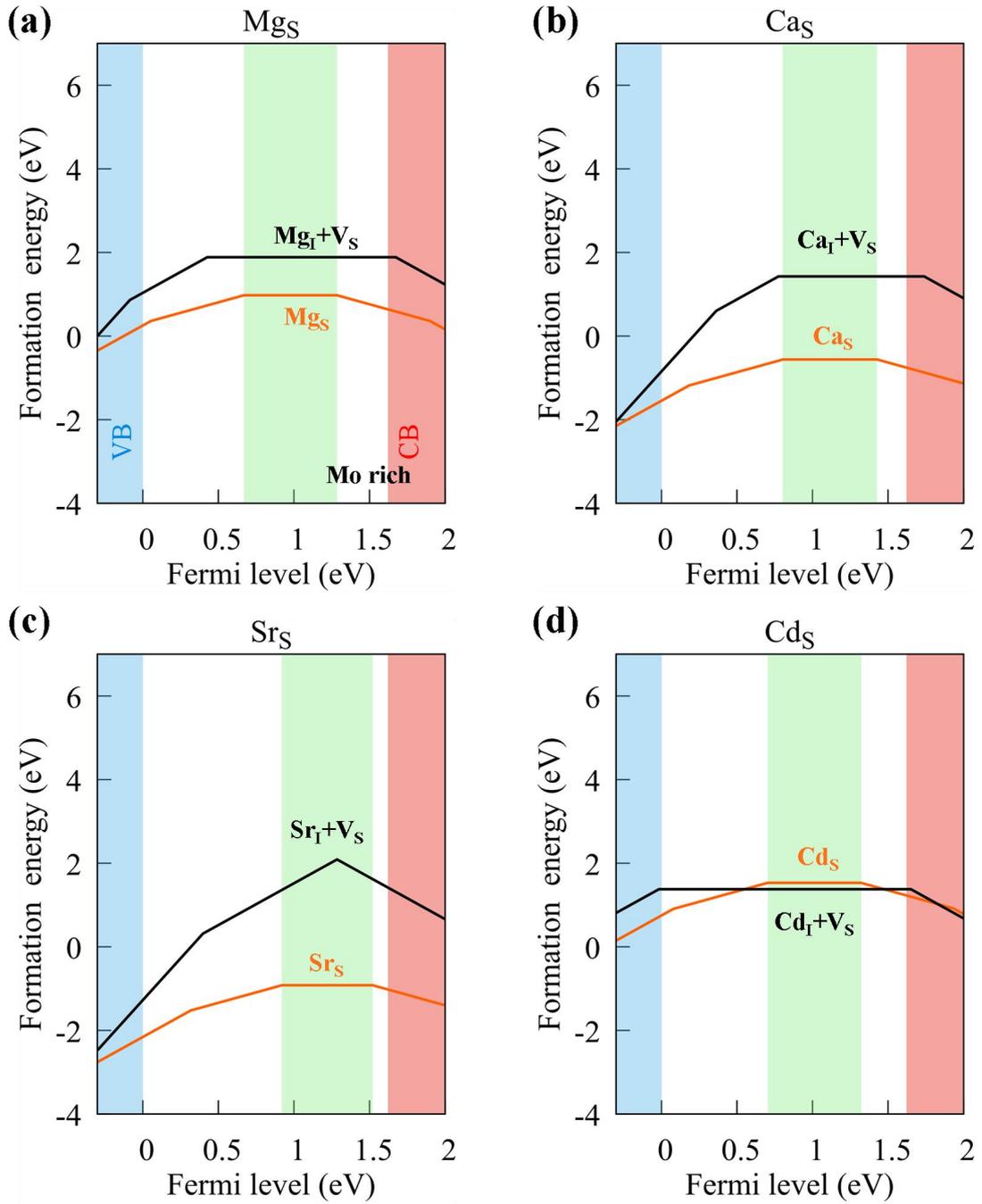

FIG. S6. Formation energies of $M_X$ defects (M = Mg, Ca, Sr, Cd; X = S) in the 6×6×1 monolayer of $MoS_2$. The formation energies of $M_X$ defects are lower than the combined energies of two independent defects ($M_I+V_X$), indicating that the formation of $M_X$ defects is energetically favored over the separate formation of $M_I$ and $V_X$ defects.

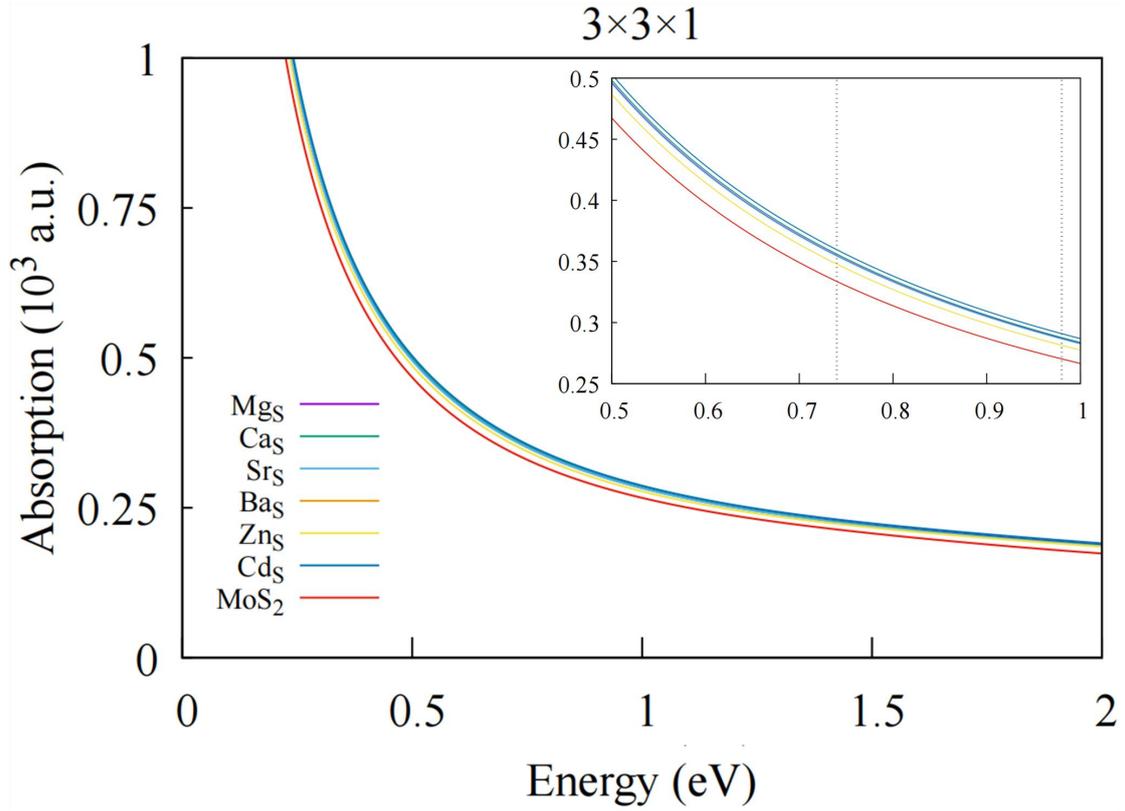

FIG. S7. Absorption spectra of monolayer $MoS_2$ with $M_X$ defects calculated for the 3×3×1 supercell. The red curve corresponds to pristine $MoS_2$; colored curves represent $M_X$-defective systems. Spectra are polarization-averaged along the x-and y-crystallographic directions. Defect-induced absorption features in the 0.740-0.984 eV range highlight mid-gap states critical for telecom-wavelength optical transitions (see region between gray lines in the inset).